\documentclass[fleqn,usenatbib]{mnras}

\usepackage{newtxtext,newtxmath}

\usepackage{threeparttable}
\usepackage[T1]{fontenc}
\usepackage{anyfontsize}
\usepackage{subfig}
\usepackage{verbatim}
\usepackage{caption}
\usepackage{subfig}

\DeclareRobustCommand{\VAN}[3]{#2}
\let\VANthebibliography\thebibliography
\def\thebibliography{\DeclareRobustCommand{\VAN}[3]{##3}\VANthebibliography}

\usepackage{graphicx}	
\usepackage{amsmath}	

\title[Clouds and Star Formation in M31]{Cloud Properties and Star Formation in M31}

\author[Armijos-Abenda\~no et al.]{J. Armijos-Abenda\~no,$^{1,2}$\thanks{E-mail: jarmijos090@gmail.com (JAA)}
	S. A. Eales,$^{1}$
	M. W. L. Smith$^{1}$
	\\
    $^{1}$School of Physics and Astronomy, Cardiff University, The Parade, Cardiff, CF24 3AA, UK\\
	$^{2}$Observatorio Astron\'omico de Quito, Observatorio Astron\'omico Nacional, Escuela Polit\'ecnica Nacional, 170403, Quito, Ecuador\\
}

\date{Accepted XXX. Received YYY; in original form ZZZ}

\pubyear{2015}

\begin{document}
	\label{firstpage}
	\pagerange{\pageref{firstpage}--\pageref{lastpage}}
	\maketitle
	
	\begin{abstract}
		We present a catalogue of 453 molecular clouds in M31 extracted from CO J=1-0 data observed with CARMA using a dendrogram. 
		Our clouds have the mean values of 2.8 km s$^{-1}$, 22.1 pc and 10$^{5.2}$ M$_\odot$ for the velocity dispersion, radius and mass, respectively. The velocity dispersion shows a weak anti-correlation with the galactocentric radius. The clouds in M31 show mean and median values of 2.0 and 1.4, respectively, for their virial parameters, indicating that most of them are gravitationally bound.
		Our dendrogram analysis identifies 35 sources with multiple velocity components, which we classify as molecular cloud complexes.
		We study the size-velocity dispersion and size-mass relationships for the clouds in M31, finding the slopes of 0.43$\pm$0.05 and 1.36$\pm$0.06 for the former and the latter, respectively. Our size-velocity dispersion relationship agrees with those of Milky Way (MW) and M31 clouds. The slope of our size-mass relationship is shallower than those in clouds and cloud complexes of the MW.
		We find offsets between the isosurfaces of the clouds and star formation rate (SFR) peaks in M31, supporting the scenario where the evolutionary state of individual sources plays a role in the Kennicutt-Schmidt (KS) law at parsec scales. We find a slope of 0.66$\pm$0.07 for the KS law, which is slightly lower than the values of $\sim$0.8 for MW clouds.
	\end{abstract}
	
	
	
	
	\section{Introduction}
	
	The proximity \citep[780 kpc,][]{2005MNRAS.356..979M} of M31 makes this galaxy an excellent laboratory for the study of the physical properties of molecular clouds and star-forming regions. 
	The molecular gas in M31 has been studied before using interferometers and also single-dish telescopes \citep{1998ApJ...499..227L,1999A&A...351.1087L,2006A&A...453..459N,2007ApJ...654..240R}. \cite{2019A&A...625A.148D} identified 12 molecular clumps towards the circumnuclear region of M31 using CO J=1-0 observations taken with the IRAM interferometer at $\sim$13 pc resolution. They found that the clumps are unbound, but this result was biased by their velocity dispersion measurements limited by the velocity resolution of $\sim$5 km s$^{-1}$. 
	Several studies have focused on comparing the physical properties of the clouds of M31 with those of the Milky Way.
	Usin BIMA observations, \cite{2000immm.proc...37S} found that six molecular cloud complexes (MCCs) in the northeastern spiral arm of M31 show a size-velocity dispersion relationship (Larson's first law) and the virial parameter ($\alpha_{\rm vir}$, the ratio between the virial mass and the source mass) comparable to those of the Milky Way. Combining interferometric and single dish observations, \cite{1998ApJ...499..227L} studied an extended cloud located at $\sim$2 kpc from the M31 centre, finding that the surface density derived from the CO emission is about a factor of 10 too low for the gas to be gravitationally bound.
	Another study by \cite{2007ApJ...654..240R} found that giant molecular clouds (GMCs) of M31 show a mean $\alpha_{\rm vir}$ of $\sim$2 consistent with that of the Milky Way. In contrast, higher values of $\alpha_{\rm vir}$ within $\sim$5-10 have been found for M31 clouds \citep{2019ApJ...883....2S}, supporting the idea that the clouds are unbound. Thus, the similarity of Milky Way clouds to those of M31 is quite controversial.
	
	Recently, \cite{2024ApJ...966..193L} discovered that 43\% of 163 $^{12}$CO GMCs in M31 are gravitationally bound. Their sample of GMCs was observed with the Submillimeter Array and selected from a catalogue of GMCs or associations of GMCs identified using the 250 $\mu$m emission. The study by \cite{2024ApJ...966..193L} supports again the idea that many properties of the clouds in M31 are very similar to those of the Milky Way. This similarity is quite striking since both galaxies have different merger histories that could have affected the formation and evolution of clouds \citep{2007ApJ...662..322H,2024ApJ...966..193L}. 
	We believe it would be interesting to study if the relationships between the physical parameters of M31 clouds holds when source extraction is performed directly from 3D data of a larger observed area, which would allow to obtain a greater number of objects existing along different lines of sight in M31 \citep{2009ApJ...705.1395C} and likely a wide distribution of the physical properties. In addition, the cloud properties can be used to infer information about the environment around them since the properties of the host galaxy such as radiation field, feedback mechanisms, ISM pressure, metallicity, etc., can regulate the formation and evolution of molecular clouds  \citep{2019ApJ...883....2S,2020SSRv..216...76B}.
	
	The formation of stars is a key element in the evolution of galaxies. It is known that the surface density of the star formation rate ($\Sigma_{\rm SFR}$) is related to the gas surface density ($\Sigma_{\rm gas}$) in the form of $\Sigma_{\rm SFR}\propto\Sigma_{\rm gas}^{\rm 1.4}$, which is called the Kennicutt-Schmidt law \citep{1998ApJ...498..541K}. \cite{2008AJ....136.2846B} found that there is a relationship of the type $\Sigma_{\rm SFR}$$\propto$$\Sigma_{\rm H_2}^{\rm N}$ with the slope N=1.0$\pm$0.2 for seven spiral galaxies on sub-kpc scales. This slope was also obtained for the star formation rate versus mass relationship for Milky Way clouds \citep{2012ApJ...745..190L}.
	\cite{2013ApJ...769...55F} discovered the $\Sigma_{\rm SFR}$$\propto$$\Sigma_{\rm H_2}^{\rm N}$ relationship with the slope N=0.60$\pm$0.01 in M31 at sub-kpc scales, implying that the Kennicutt-Schmidt relationship is not superlinear in M31.
	
	This work aims to create a sample of GMCs for M31 by applying a dendrogram to CARMA data in the position-position-velocity space, which will create the largest cloud sample for M31 so far. Then, we will determine the physical properties of our sample and study the size-velocity dispersion and size-mass relationships, which will allow rechecking the similarity between the M31 clouds and those of the Milky Way that has been proposed before \citep{2000immm.proc...37S,2007ApJ...654..240R,2024ApJ...966..193L}, but now using a larger number of objects extracted from 3D data. 
    We will also investigate if the Kennicutt-Schmidt law found by \cite{2013ApJ...769...55F} holds when determined based on a cloud-by-cloud analysis, which is different from their method based on a pixel-by-pixel analysis. 
	Our work is organized as follows. We describe data reduction in Section \ref{Reduction}, while the extraction of GMCs using a dendrogram analysis is explained in Section \ref{extraction}. The determination of the source properties and the analysis of multiple gas components along the line of sight of several clouds are presented in Sections \ref{properties} and \ref{COspectra_sec}, respectively. We study the size-velocity dispersion relationship in Section \ref{Larson_first} and the size-mass relationship in Section \ref{Sizevsmass}. The Kennicutt-Schmidt law for our M31 sample is presented in Section \ref{KSlaw}.
	Finally, the conclusions are given in Section \ref{Conclusions}.
	
	\section{Data reduction}\label{Reduction}
	
\begin{figure}
		\centering
		\includegraphics[trim={3cm 0 1cm 1cm},clip,width=0.5\textwidth]{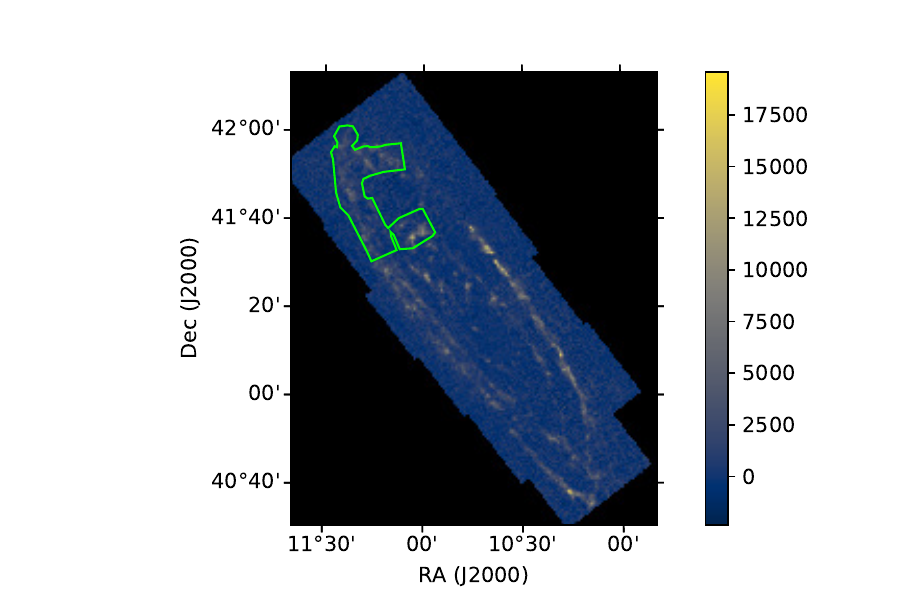}
		\caption{Integrated intensity map of the CO J=1-0 line emission of M31 in units of K m s$^{-1}$ \citep{2006A&A...453..459N}. The green line indicates the region mapped with CARMA.}
		\label{CO_map}
	\end{figure}

	We used observations available in the CARMA database\footnote{http://carma-server.ncsa.uiuc.edu:8181/asp/carmaQuery.cgi}. The observations were carried out between 2011 and 2014 (Projects C0803, C0957, C1200, and C1126) using the compact D and E configurations. The CO J=1-0 line at 115.271 GHz was observed with one spectral window of 255 channels, providing a spectral resolution of 0.727 km s$^{-1}$. The CARMA observations covered parts of the 5 kpc and 10 kpc rings of M31 (see Figure \ref{CO_map}).
	
	We carried out the data reduction following the standard reduction procedures using MIRIAD \citep{1995ASPC...77..433S}.
	The passband calibration was done using bright sources, usually 3C84 or 3C454.3. Data with bad baselines, atennas, and/or time-ranges were flagged. Uranus or the compact HII region MWC349 was used as flux calibrator. The calibrated visibilities were imaged in MIRIAD with the task MOSSDI that uses the CLEAN algorithm of Steer \citep{1984A&A...137..159S}. Then, we obtained a cleaned data cube of the CO J=1-0 transition with the synthesized beam of 7.0$\arcsec\times$4.1$\arcsec$ (27 pc $\times$ 16 pc at the distance of M31).
	
	Using MIRIAD, we also combined the CARMA data with CO J=1-0 IRAM 30m observations of M31 \citep{2006A&A...453..459N} for correcting the missing flux due to extended spatial scales filtered out by the interferometric observations. The spectral resolution of the CARMA data was degraded to that of 2.5 km s$^{-1}$ of the IRAM 30m data before the combination. 
	Thus, we obtained a combined data cube of the CO J=1-0 transition with noise levels of $\sim$200-300 mK. 
	\cite{2016AJ....151...34C} applied a similar method to the same data set for addressing the problem of the missing flux. The data cube has spatial pixels of 2\arcsec$\times$2\arcsec and spectral pixels of 2.5 km s$^{-1}$.
	
	\section{Results}
	\subsection{Cloud extraction}\label{extraction}
	
\begin{figure*}
		\centering
		\includegraphics[trim={1.1cm 1.1cm 2cm 2cm},clip,width=20cm]{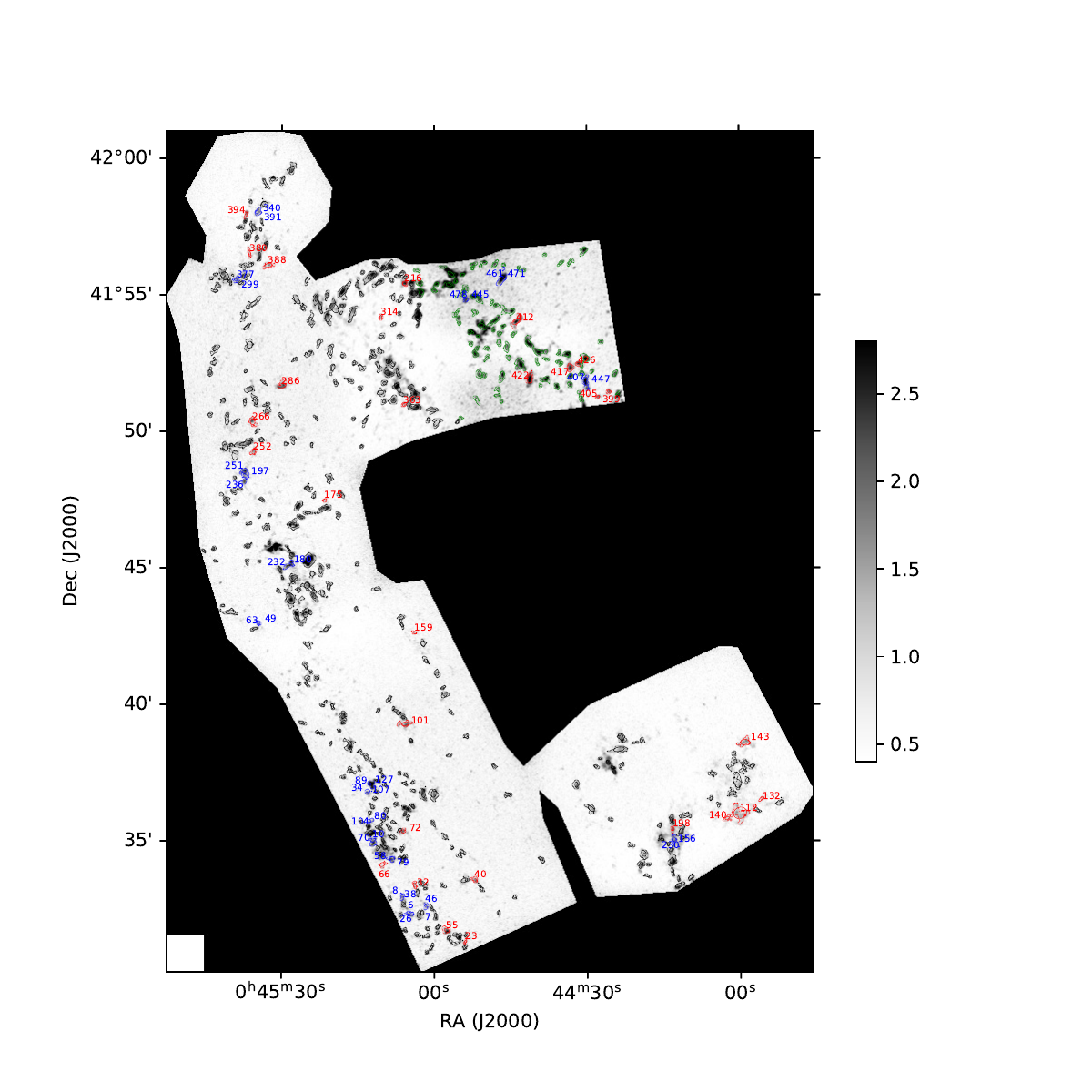}
		\caption{Peak intensity map of the CO J=1-0 transition in units of Kelvin. The red, black, blue, and green contours show the boundaries of 488 sources identified using a dendrogram. The extent of each source is outlined using a 3$\sigma$ threshold. This figure must be enlarged considerably to visualize details of the identified objects. A $\sigma$ of 300 mK was used for identifying sources in the regions where the majority of the sources are outlined with green contours, while a $\sigma$ of 230 mK was used for identifying sources elsewhere. The blue contours outline 35 sources with multiple velocity components identified in our dendrogram analysis. The spectra of 11 of these multiple velocity component sources and of 29 randomly selected sources (outlined with red contours) are shown in Figure \ref{COspectra}. All the sources with multiple components and the 29 sources outlined with red contours are labeled with a number following the numbering in Table \ref{cloud_catalogue}. The best-fit ellipses that follow the source isosurfaces are also indicated.}\label{Identified_clouds}
	\end{figure*}

	To study the cloud properties in M31 we extracted clouds using the Python package ASTRODENDRO 0.2.0 \citep{2008ApJ...679.1338R}, which decomposes data sets into hierarchical structures called leaves, branches, and trunks. For the computation of the dendrogram, we specified a minimum intensity value of a pixel of 3$\sigma$, a minimum significance value of 2$\sigma$ as the minimum difference in the peak intensity between neighbouring structures, and a minimum number of pixels (min\_npix) equal to the number of pixels (21) contained in 2.5 the telescope-beam solid angle. The min\_npix value is the minimum value that a leaf has to have to be considered an independent identity. The min\_npix of 21 is a conservative value. Smaller min\_npix values lead to the identification of small spurious sources, while on the other hand, a min\_npix value greater than 21 leads to real leaves merging with a branch or another leaf.
	We applied the dendrogram analysis to the combined CO J=1-0 data cube in position-position-velocity space. Thus, we identified 488 sources indicated in Figure \ref{Identified_clouds}. A region of the CO J=1-0 data cube has noise levels greater than in the rest of the observed areas due to differences in the amount of available visibility data.
	The noise levels in both high and low noise regions are average values measured across three spatial areas of the data cube. Importantly, the noise levels remain relatively stable, regardless of the aperture size used for measurement. We calculated the noise levels using line-free channels with a resolution of 2.5 km s$^{-1}$. We used $\sigma$=300 mK for the area with greater noise levels, while $\sigma$=230 mK was used for the remaining regions.
	
\begin{figure*}
		\centering
		\includegraphics[trim={3.5cm 0 2.5cm 0},clip,width=18cm]{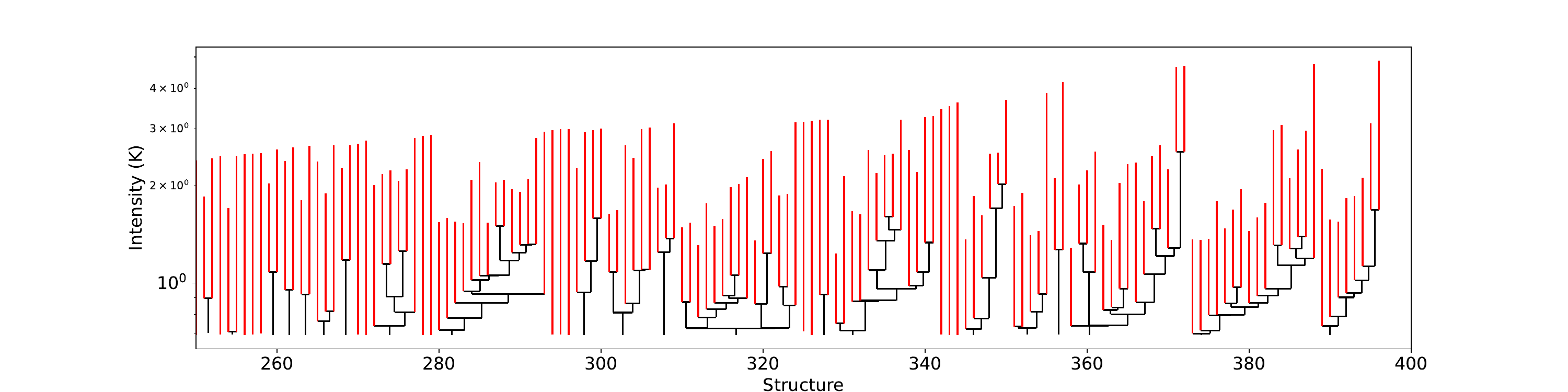}
		\caption{Part of the dendrogram of the CO J=1-0 emission. The leaves highlighted in red are considered clouds.}
		\label{dendrotree}
	\end{figure*}

\begin{figure*}
    \centering
    \subfloat{\includegraphics[width=.8\linewidth]{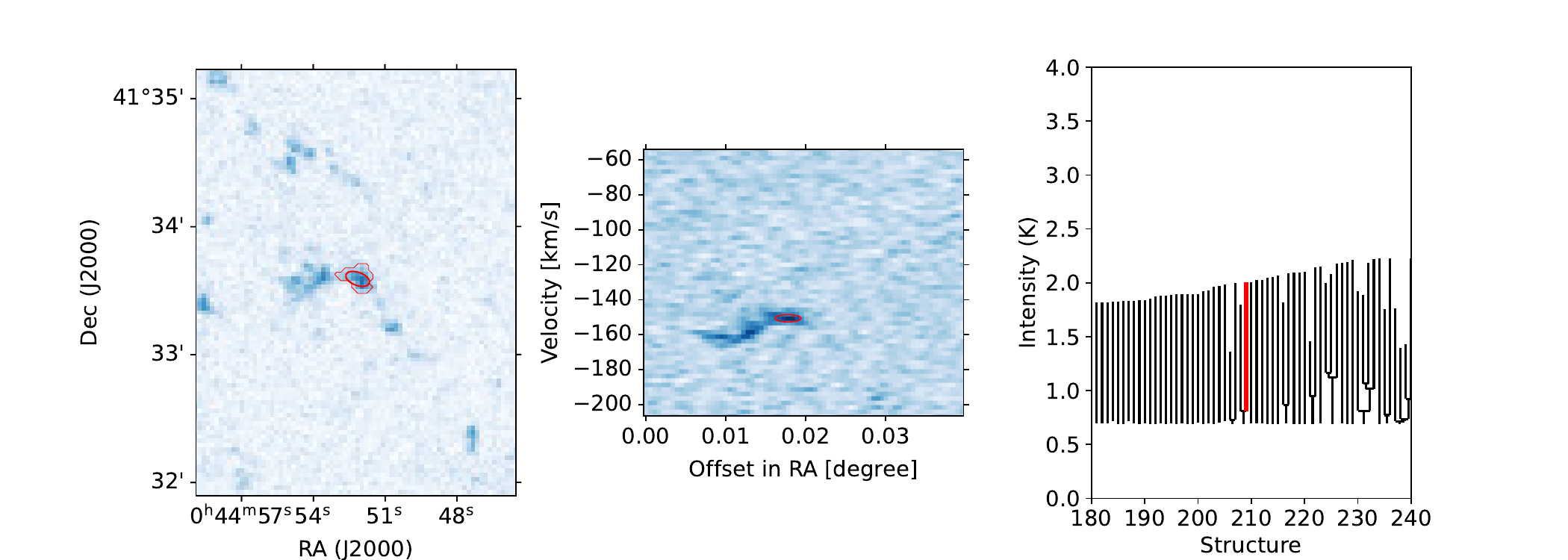}}\hfill
    \subfloat{\includegraphics[width=.8\linewidth]{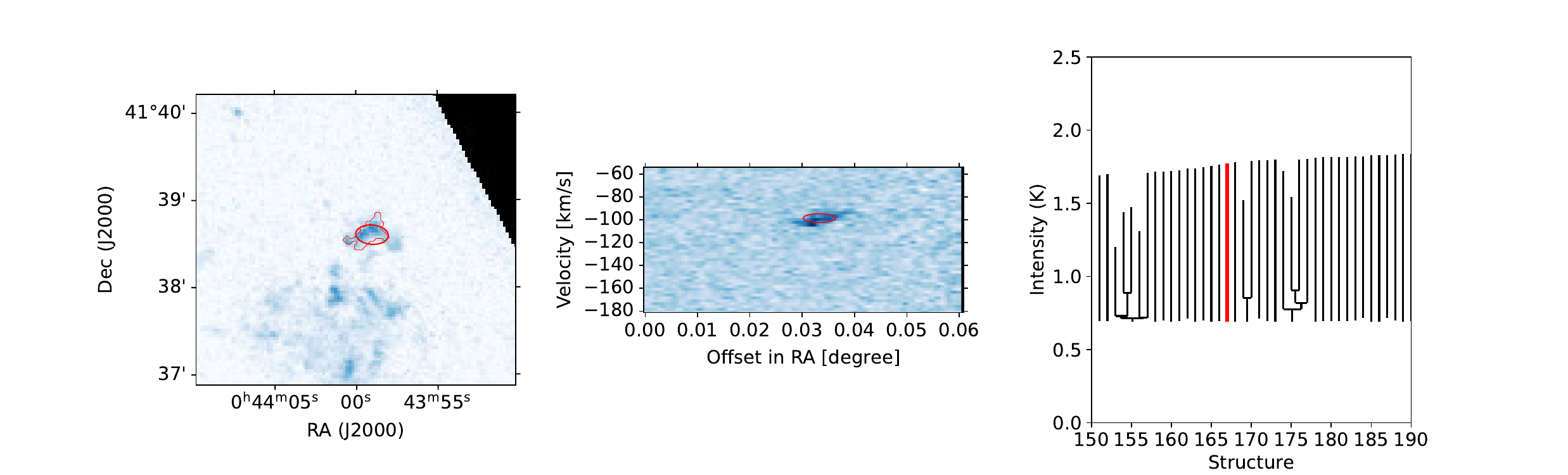}}\hfill
    \subfloat{\includegraphics[width=.8\linewidth]{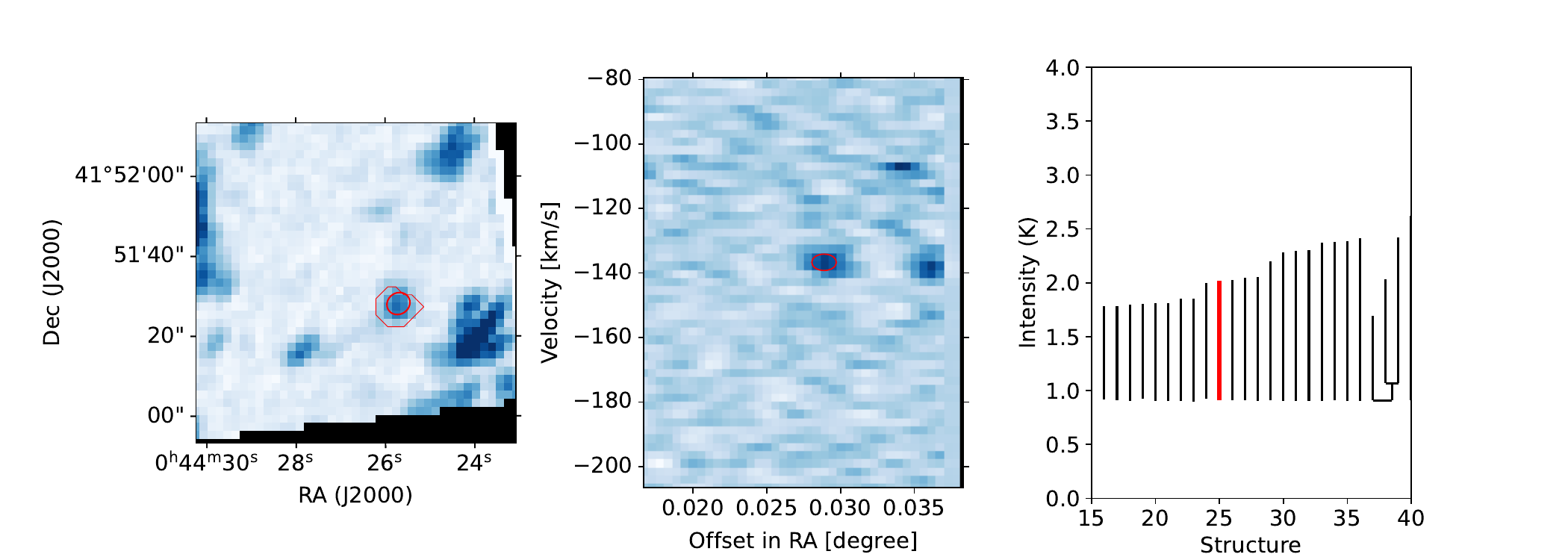}}
	\caption{\textbf{Top:} the red ellipse is the best-fit to cloud 40 on the peak intensity map of CO J=1-0 (left panel) and on a position-velocity map (center panel) identified by the dendrogram. The position-velocity map is extracted along a path horizontal to the X axis, passing through the center of the ellipse in the left panel. The red contour corresponds to 3$\sigma$ in the left panel. The right panel shows a portion of the dendrogram, with cloud 40 highlighted in red. \textbf{Medium:} the same as in the top but for cloud 143. \textbf{Bottom:} the same as in the top but for cloud 399.}
	\label{example_clouds}
    \end{figure*}

	Our research focuses on the study of dendrogram leaves, which are considered molecular clouds. Towards several clouds, we observe more than one velocity component, categorizing them as molecular complexes. In our analysis, 35 out of the 488 sources showed multiple velocity components. 
	No threshold was used when determining the velocity dispersion of the identified M31 clouds; however, we exclude sources from our analysis when the velocity dispersion was influenced by instrumental resolution (refer to Section \ref{properties}). Figure \ref{dendrotree} shows a portion of the dendrogram for the CO(J=1-0) emission. Figure \ref{example_clouds} provides examples of the dendrogram extraction of three specific clouds.
	
	\subsection{Cloud properties}\label{properties}
	
	The dendrogram provides the velocity dispersion $\sigma_{\rm v}$, as well as the major axis radius ($\sigma_{\rm maj}$) and minor axis radius ($\sigma_{\rm min}$) of the best-fit to the identified cloud. Following \cite{2006PASP..118..590R}, we calculated the deconvolved one-dimensional size of a cloud by:
	
	\begin{equation}\label{ecuacion1}
		\sigma_r^d=\sqrt{ \left(\sigma_{\rm maj}^2-\frac{\Theta_{\rm major}^2}{8\,\ln(2)}   \right)^{1/2} \left( \sigma_{\rm min}^2-\frac{\Theta_{\rm minor}^2}{8\,\ln(2)}  \right)^{1/2}    }
	\end{equation}
	
	where $\Theta_{\rm major}$ and $\Theta_{\rm minor}$ are the major and minor axes, respectively, of the telescope beam. Then, we multiplied the $\sigma_r^d$ value by 1.91 to obtain the spherical radius $r_{\rm c}$ of a cloud \citep{2006PASP..118..590R}.
	
	The deconvolved velocity dispersion is calculated by:
	
	\begin{equation}\label{ecuacion2}
		\sigma_v^d=\sqrt{\sigma_{\rm v}^2-\frac{\Delta_v^2}{8\,\ln(2)}}
	\end{equation}
	
	where $\Delta_v$ is the the channel width.
	
	Table \ref{cloud_catalogue} gives the cloud number, the Right Ascension and Declination in epoch J2000 coordinates, the cloud radius $r_{\rm c}$, and the velocity dispersion $\sigma_{\rm v}^d$ obtained for the identified clouds in M31. We do not provide the values of $r_{\rm c}$ and/or $\sigma_v^d$ in Table \ref{cloud_catalogue} when equations \ref{ecuacion1} and/or \ref{ecuacion2} yield indeterminate values. Out of the 453 clouds in our catalog (excluding the molecular cloud complexes), 118 have indeterminate values for $r_{\rm c}$ and/or $\sigma_v^d$. In the analysis that follows, we have considered only 335 sources, unless stated otherwise. 
	
	The integrated flux of the CO J=1-0 emission for each source is calculated by the dendrogram \citep{2006PASP..118..590R} using the following equation:
	
	\begin{equation}
		F_{\rm CO}=\sum_{i} T_i \, A_{\rm pix} \, \Delta_{\rm v}
	\end{equation}
	
	where T$_{\rm i}$ is the intensity in a given pixel measured in Kelvin, $A_{\rm pix}$ is the pixel area in arcsec$^2$, and $\Delta_{\rm v}$ is again the channel width in km s$^{-1}$.
	
	In addition, we have estimated the mass ($M_{\rm c}$) for the sources in M31 (listed in Table \ref{cloud_catalogue}) using the above CO flux in the following expression:
	
	\begin{equation}
		M_{\rm c} = 1.36\left(\frac{\pi}{180\times3600}\right)^2 F_{\rm CO} d^2 X_{\rm CO} m_{\rm H_2}
	\end{equation}
	
	where $d$ is the distance to M31 equal to 785 kpc \citep{2005MNRAS.356..979M}, $X_{\rm CO}$=1.9$\times$10$^{20}$ cm$^{-2}$ [K km s$^{-1}$]$^{-1}$ is the CO-to-H$_{\rm 2}$ conversion factor \citep{1996A&A...308L..21S}, and $m_{\rm H_2}$ is the mass of a hydrogen molecule in kg. The factor of 1.36 is included in the above equation to account for helium.
    The mass of our M31 clouds is not corrected by metallicity since there does not appear to be a clear gradient in metallicity with the galactocentric radius in M31 \citep{2022MNRAS.517.2343B}.
	
	To estimate errors of the above properties we use the bootstrap method following \cite{2006PASP..118..590R}. 
	We generate a simulated cloud by replacing the flux in all the spectral pixels of an original cloud randomly. This was done for each cloud 100 times, allowing for repeated draws.
	Then, we recalculated the cloud properties using a dendrogram for each of those trial clouds. 
	The error of each property is the standard deviation of the recalculated quantities for each trial cloud.
	The final error for each cloud property is obtained rescaling the standard deviation by the root squared of the oversampling rate ($\sqrt{\frac{1.1331\Theta_{\rm maj}\Theta_{\rm min}}{A_{\rm pix}}}$, where $\Theta_{\rm maj}$ and $\Theta_{\rm min}$ are the major and minor FWHM values, respectively, of the synthesized beam) to account for the lack of independence of some pixels in each source. The errors of other cloud properties derived from the $\sigma_{\rm maj}$, $\sigma_{\rm min}$, $\sigma_{\rm v}^d$, and $F_{\rm CO}$ values were calculated through error propagation. We calculate a large error in the source radius when the $\sigma_{\rm maj}\sim\frac{\Theta_{\rm major}}{2.35}$ and/or the $\sigma_{\rm min}\sim\frac{\Theta_{\rm minor}}{2.35}$, while a large error in the velocity dispersion is determined when the $\sigma_{\rm v}\sim\frac{\Delta_{\rm v}}{2.35}$, that is, when the $\sigma_{\rm maj}$, $\sigma_{\rm min}$ and $\sigma_{\rm v}$ values are close to their resolution limits.
	
\begin{figure}
	\includegraphics[trim={1cm 0 0 0},width=9cm]{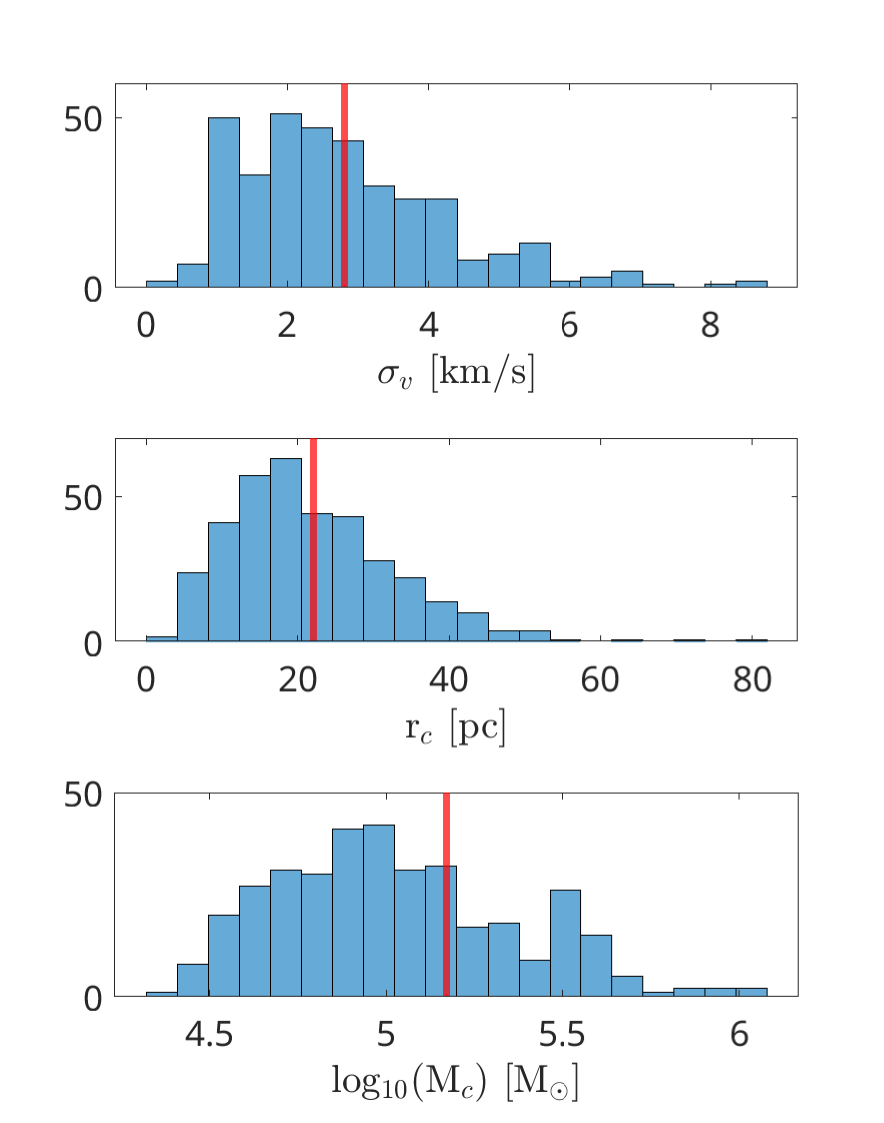}
	\caption{Histograms showing the distribution of the $\sigma_{\rm v}^d$ (top panel), r$_{\rm c}$ (medium panel) and $M_{\rm c}$ (bottom panel). Red lines are the mean values of 2.8 km s$^{-1}$, 22.1 pc and 5.2 M$_{\odot}$ for $\sigma_{\rm v}^d$, $r_{\rm c}$, and log10(M$_{\rm c}$), respectively.}
	\label{histograms}
	\end{figure}

	The distribution of the $\sigma_{\rm v}^d$, $r_{\rm c}$ and $M_{\rm c}$ values is given in Figure \ref{histograms}, where we indicate the mean values of 2.8 km s$^{-1}$ for the $\sigma_{\rm v}^d$, of 22.1 pc for the $r_{\rm c}$, and of 5.2 for the $\log_{10} (M_{\rm c})$. 
	
	\begin{figure}
	\includegraphics[trim={1.3cm 0 1cm 0},width=9cm]{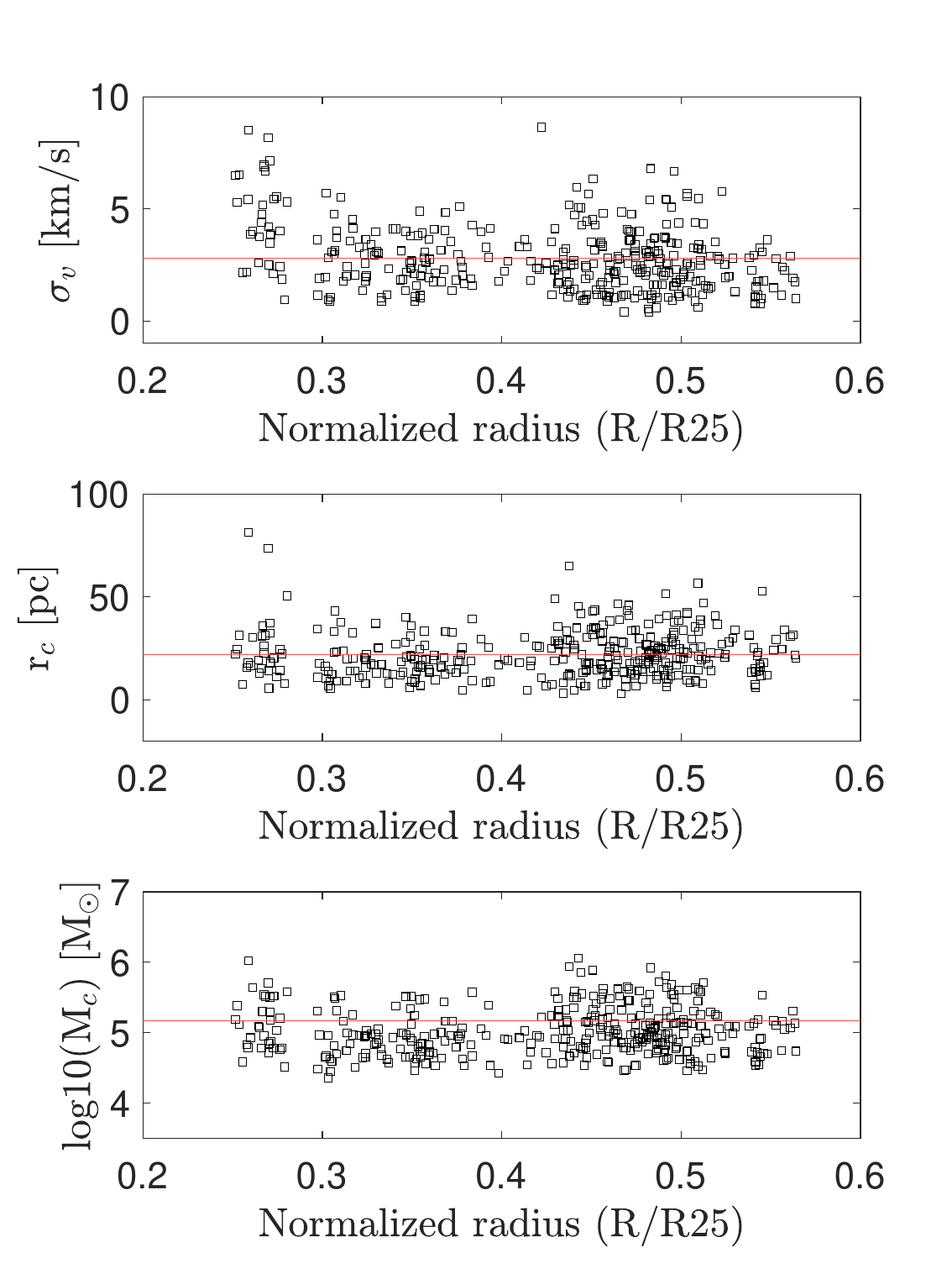}
	\caption{Normalized galactocentric radial distribution of $\sigma_{\rm v}^d$ (top panel), r$_{\rm c}$ (medium panel), and cloud mass (bottom panel). The red line in each panel shows the mean value.} 
	\label{radial_distribution}
	\end{figure}
	
	We also present the values of these parameters as a function of the galactocentric radius ($R$) normalized by $R_{\rm 25}$ (the optical radius equal to 21.55 kpc) in Figure \ref{radial_distribution}.
	We give in Table \ref{Correlation-coef} the Pearson correlation coefficients and p-values for the values given in Figure \ref{radial_distribution}, which show that there is no correlations of R/R25 versus $r_{\rm c}$ and R/R25 versus $\log_{10}(M_{\rm c})$. At the same time, there is a weak inverse correlation between R/R25 and $\sigma_{\rm v}^d$.
	
	\begin{table}
	\centering
	\caption{Pearson correlation coefficients and p-values.}
	\label{Correlation-coef}
	\begin{tabular}{llcc} 
			\hline
			Data set & Correlation coefficient & p-value \\
			\hline
			$\sigma_{\rm v}^d$ versus R/R25     &  -0.29  & 8.72$\times$10$^{-8}$ \\
			r$_{\rm c}$ versus R/R25            &   0.08 & 0.14 \\
			$\log_{10}(M_{\rm c})$ versus R/R25 &   0.02 & 0.77 \\
			\hline
	\end{tabular}
	\end{table}	
	
\begin{figure}
		\subfloat{\includegraphics[trim={0 0 1cm 0},clip,width=8cm]{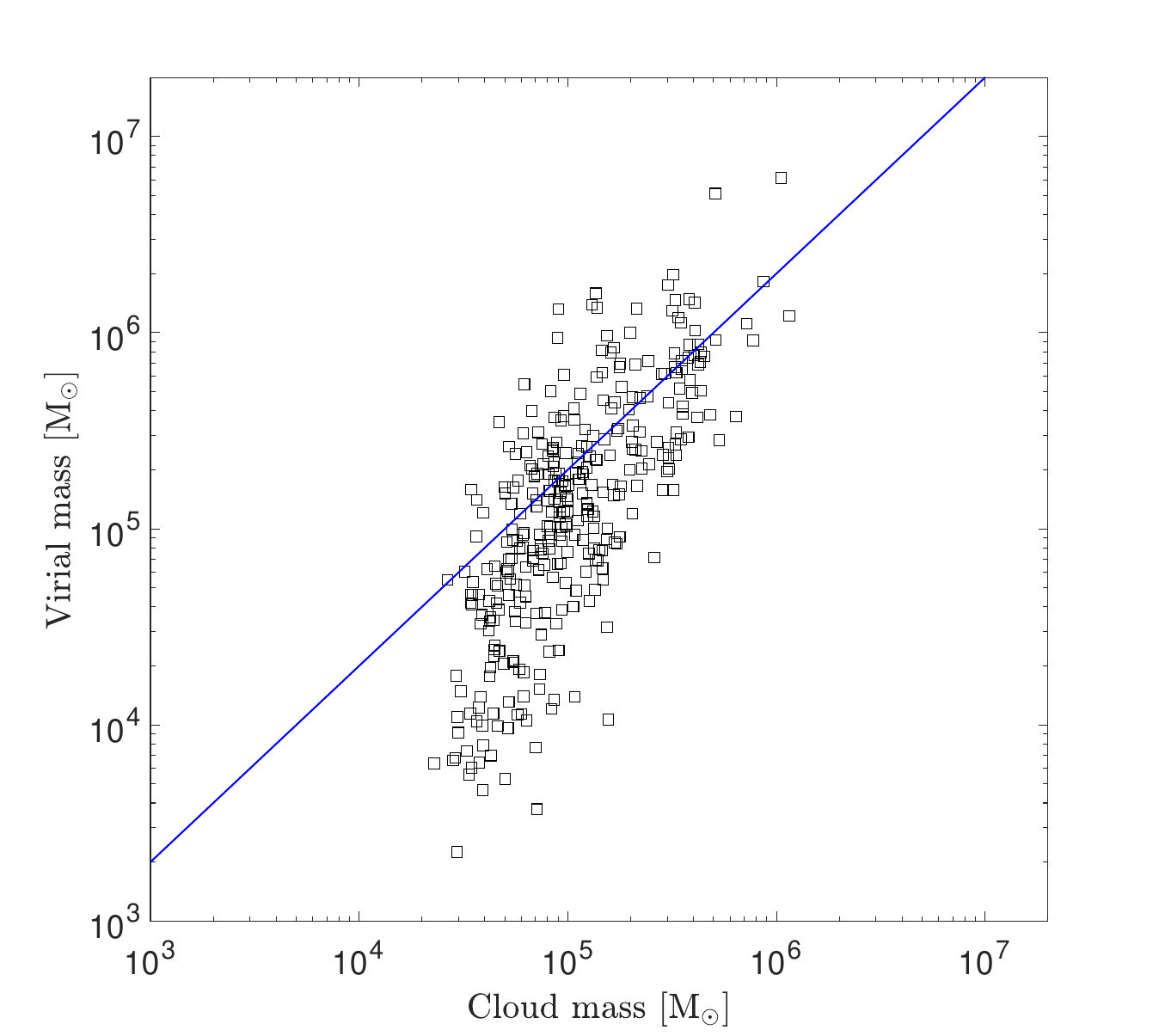}}\hfil
	    \subfloat{\includegraphics[trim={0 0 0.5cm 0},clip,width=8.5cm]{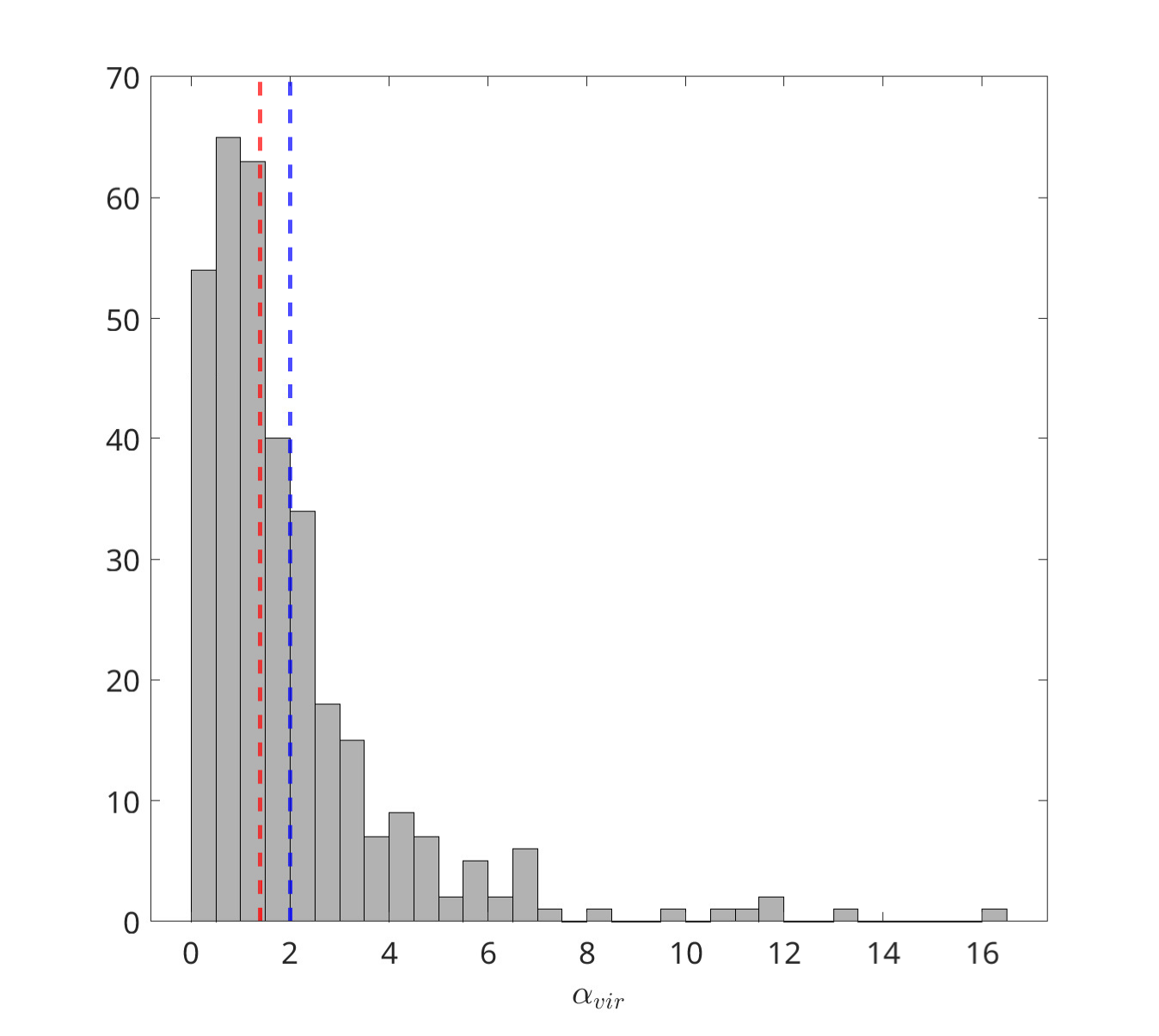}}
		\caption{\textbf{Top panel:} the $M_{c}$ values as a function of the $M_{\rm vir}$ values for the clouds of M31. The blue line is $M_{\rm vir}$=2$M_{\rm c}$. 114 of the sources (336) shown in this figure have values above the blue line. \textbf{Bottom panel:} distribution of the $\alpha_{\rm vir}$ parameter for the sources of M31 with a median value of 1.4 (dashed red line). $\alpha_{\rm vir}$=2 is indicated with the dashed-blue line (the threshold between gravitationally bound and unbound objects).} 
		\label{Mh2Mvir}
	\end{figure}

    Following \cite{2019MNRAS.483.4291C}, we also determined the virial mass of the sources as:
	
	\begin{align}
		M_{\rm vir} &= 1040\,\sigma_v^2 r_{\rm c},
	\end{align}
	
	where we use the $\sigma_{\rm v}^d$ values for $\sigma_{\rm v}$. The virial mass for our source catalogue is given in Table \ref{cloud_catalogue}. 
	The top panel of Figure \ref{Mh2Mvir} shows the $M_{\rm vir}$ values versus the $M_{\rm c}$ values for our M31 clouds. We calculated the virial ratio $\alpha_{\rm vir}=\frac{M_{\rm vir}}{M_{\rm c}}$ \citep{1992ApJ...395..140B} for these sources as well. Bound clouds are characterized by $\alpha_{\rm vir}\leqslant 2$, while unbound clouds are characterized by $\alpha_{\rm vir}> 2$.
	We find that 114 out of 336 clouds in the top panel of Figure \ref{Mh2Mvir} have values above the line M$_{\rm vir}$=2$M_{\rm c}$. This indicates that 66\% of our clouds are gravitationally bound. The bottom panel of Figure \ref{Mh2Mvir} depicts a histogram of the $\alpha_{\rm vir}$ values. We find a median value of 1.4 and a mean value of 2.0 for our $\alpha_{\rm vir}$ values. Our mean $\alpha_{\rm vir}$ value agrees with that of 2.0$\pm$0.3 found by \cite{2007ApJ...654..240R} for clouds in M31.
	However, our mean $\alpha_{\rm vir}$ is lower than the 5.6$\pm0.55$ reported for a sample of disk clouds in M31 \citep{2025MNRAS.538.2445D}. This discrepancy may arise because the resolution of their data is a factor of $\sim$3 coarser than ours, which could mean that some of their objects are cloud complexes rather than individual clouds. 
	
\begin{figure}
		\includegraphics[trim={1.0cm 0 0 0},width=8cm]{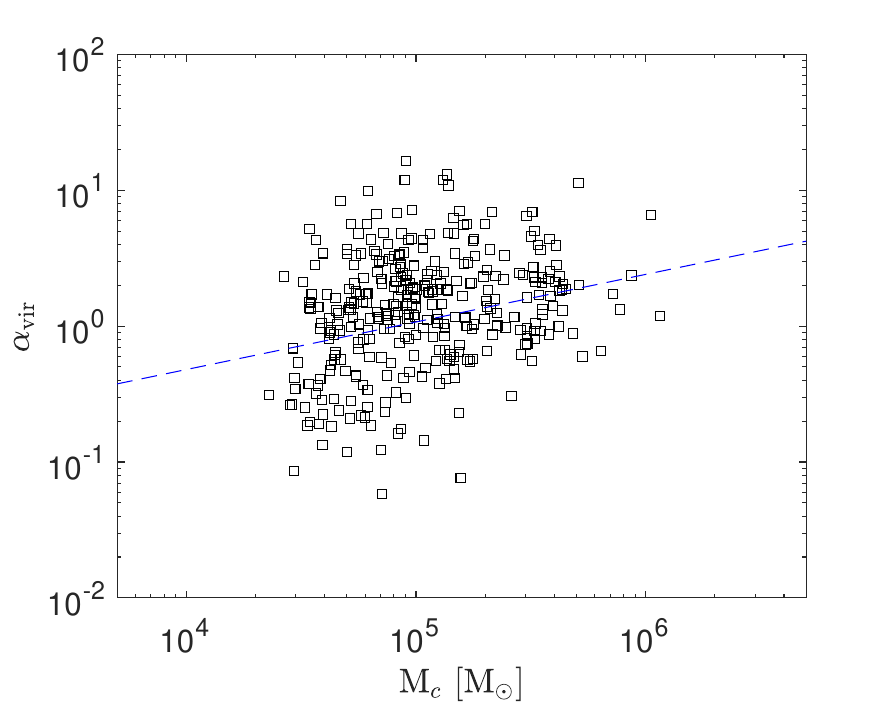}
		\caption{Mass as a function of the $\alpha_{\rm vir}$ for the sources in M31. The dash-blue line is a fit to our data (see Section \ref{properties}).}\label{massvsalpha}
	\end{figure}

	It has been proposed that pressure confinement by the diffuse ambient medium is an important driver for high $\alpha_{\rm vir}$ values found in M31 \citep{2019ApJ...883....2S}. We found that there is a weak dependence between the source mass and the $\alpha_{\rm vir}$ (see Figure \ref{massvsalpha}) in the form of $\alpha_{\rm vir}\,\propto\,M_{\rm c}^{0.34}$, which contradicts the relationship $\alpha_{\rm vir}\propto\,M^{-2/3}$ expected for sources confined by external pressure \citep{1992ApJ...395..140B}. Our best fit to our data is listed in Table \ref{alpha_mass}.
	
\begin{figure}
		\includegraphics[trim={0.5cm 0 0 0},width=9cm]{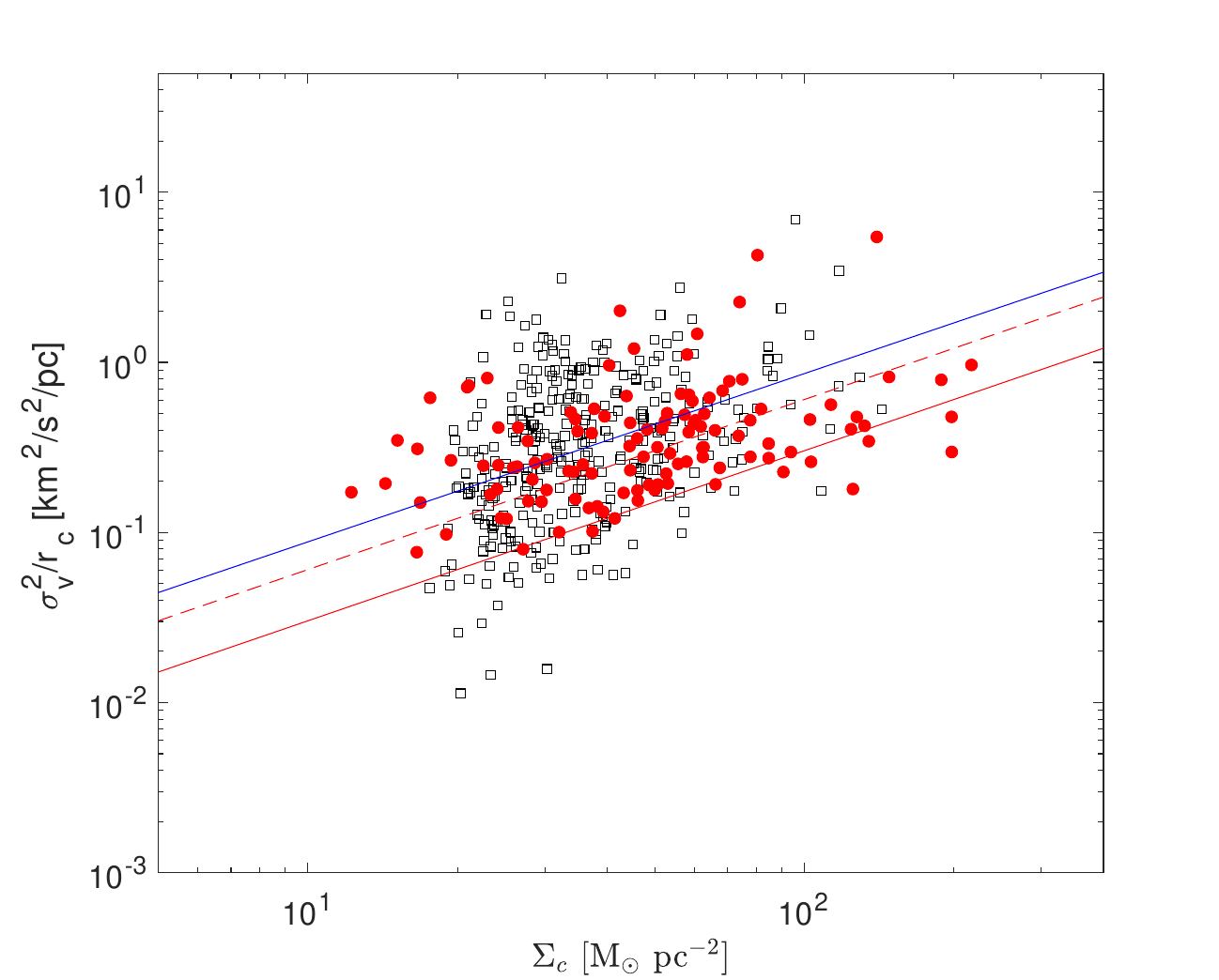}
		\caption{The values derived for our sample in M31. The threshold between bound and unbound sources is indicated with the dashed-red line, while the blue line is the best fit to our data (see Section \ref{properties}). The case of virialization is indicated with the red line. The filled red circles are data taken from a previous study \citep{2024ApJ...966..193L}.}\label{Surfacevssigrad}
	\end{figure}

	\begin{table}
		\centering
		\caption{Best fit parameters for the $\alpha_{\rm vir}=a\,M_{\rm c}^N$ relationship.}
		\label{alpha_mass}
		\begin{tabular}{lccrr} 
			\hline
			Data set & a & N  & C$_{\rm corr}$ & p-value\\
			\hline
			Our data & 10$^{-1.58\pm0.33}$ & 0.34$\pm$0.07 & 0.27 & 6.47$\times$10$^{-7}$\\
			\hline
		\end{tabular}
	\end{table}
	Considering Larson's first law we can relate the velocity dispersion and source radius with the surface density ($\Sigma_{\rm c}$) as $\frac{\sigma_{\rm v}^2}{r_{\rm c}}=\left(\frac{\pi\,G}{5}\right)\Sigma_{\rm c}$. This relationship expected in the case of virialization is shown in Figure \ref{Surfacevssigrad}, where we have also plotted our data. The $\Sigma_{\rm c}$ values are estimated as the ratio between the cloud mass over the exact area of the structure provided by our dendrogram analysis.
	We find a relationship of the form $\frac{\sigma_{\rm v}^2}{r_{\rm c}}=10^{-2.04\pm0.18}\Sigma_{\rm c}^{0.99\pm0.12}$ with a correlation coefficient of 0.40 when fitting our data given in Figure \ref{Surfacevssigrad}. The slope of 1 we found is consistent with the concept of virialization. The parameter $a=10^{-2.04}$ is just a factor of $\sim$1.5 greater than the threshold that distinguishes between bond and unbound sources, as indicated in Figure \ref{Surfacevssigrad}. This supports the conclusion that the majority of our sources are bound.
	
	Additionally, we have included data on M31 clouds from \cite{2024ApJ...966..193L} in Figure \ref{Surfacevssigrad}. We can see in this figure that our values align well with those reported by \cite{2024ApJ...966..193L}. Recently, \cite{2025ApJ...986...12L} found that most of the M31 clouds studied in \cite{2024ApJ...966..193L} are either in or near virial equilibrium, due to a combination of surface pressure and gravity. To reach this result, the authors studied the surface density versus internal pressure ($p_{\rm int}(r)=\Sigma(r)\frac{\sigma_{\rm v}^2}{r}$) profiles of the M31 clouds, demonstrating that these profiles follow the expected relationship of $p_{\rm int}\sim\Sigma^2$ found in hydrostatic equilibrium.
	
	\subsection{Sample of CO J=1-0 spectra}\label{COspectra_sec}
	
	We show a sample of 40 spectra in Figure \ref{COspectra}, which were integrated over the exact area of sources selected randomly and highlighted with red (29 sources) or blue (11 out of 35 sources with multiple velocity components) contours in Figure \ref{Identified_clouds}.
	The source number of the 40 selected sources in Figure \ref{Identified_clouds} is the same as indicated in Table \ref{cloud_catalogue}.
	The exact area (A$_{\rm exact}$) of the sources is the area of the structure in the sky and it is determined by our dendrogram analysis. This area is used to derive a factor given by $f=\frac{1}{\sqrt{\sigma_{\rm maj}\,\sigma_{\rm min}}}\sqrt{\frac{A_{\rm exact}}{\pi}}$, which in turn is used to determine the axes ($\sigma_{\rm maj}^{\rm exact}=f\sigma_{\rm maj}$ and $\sigma_{\rm min}^{\rm exact}=f\sigma_{\rm min}$) of an ellipse with an exact area of the structure with a position angle (PA) also provided by our dendrogram analysis.
	These ellipses with the exact area of the identified sources will be used in Section \ref{KSlaw}, where we will study the relationship between the star formation rate (SFR) and the source mass (derived from the CO J=1-0 line flux measured over the isosurface in our dendrogram analysis). 

	We see in Figure \ref{Identified_clouds} that the ellipses defined by the best-fit parameters follow well the boundaries of the source isosurfaces. 
	Here we only use the ellipses to extract CO J=1-0 spectra to study sources with multiple velocity components.
	The velocity of the gas component identified in our dendrogram analysis is also marked in the spectra of Figure \ref{COspectra}. This figure reveals that there are multiple velocity components along the line of sight in different sources towards M31. More than one velocity component is picked up by our dendrogram analysis along several lines of sight in our data as there are overlapped contours towards the sources outlined with blue contours in Figure \ref{Identified_clouds}. These multiple velocity component sources are marked in Table \ref{cloud_catalogue}. We find 35 sources that are in regions where there is more than one velocity component along the line of sight: 16 sources with two velocity components and 1 source with three velocity components. Of these 35 objects, 10 have indeterminate values of the deconvolved radius and/or velocity dispersion (see Table \ref{cloud_catalogue}). 
	
	\begin{figure*}
		\includegraphics[width=17cm]{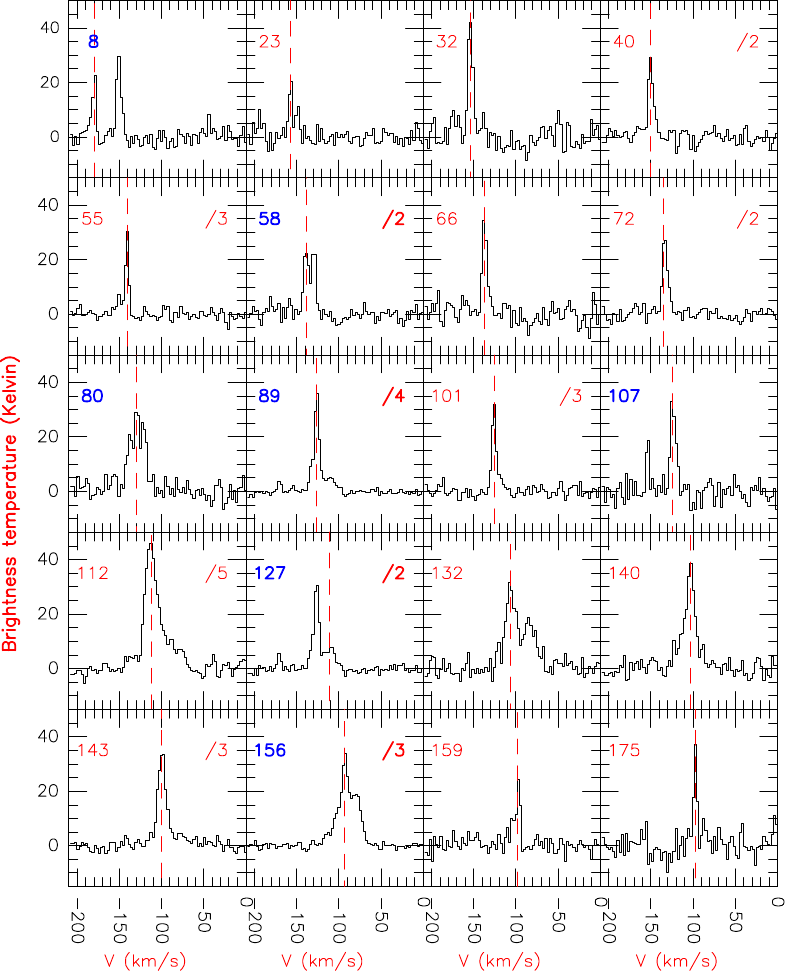}
		\caption{Spectra integrated over the exact area of forty sources identified with a dendrogram and shown with red (31 sources) and blue (11 multiple velocity component sources of several) contours in Figure \ref{Identified_clouds}. The source number (in blue for multiple velocity sources and in red for sources with only one identified velocity component by our dendrogram analysis) listed in Table \ref{cloud_catalogue} is indicated in the top-left corner of each panel. Some spectra are divided by a number (indicated on the top-right corner of each panel) for better visualization. The vertical dashed-red line in each panel indicates the velocity of the gas component identified by our dendrogram analysis.
		The next part of this figure continues below.}\label{COspectra}
	\end{figure*}
	
	\begin{figure*}
	\includegraphics[width=17cm]{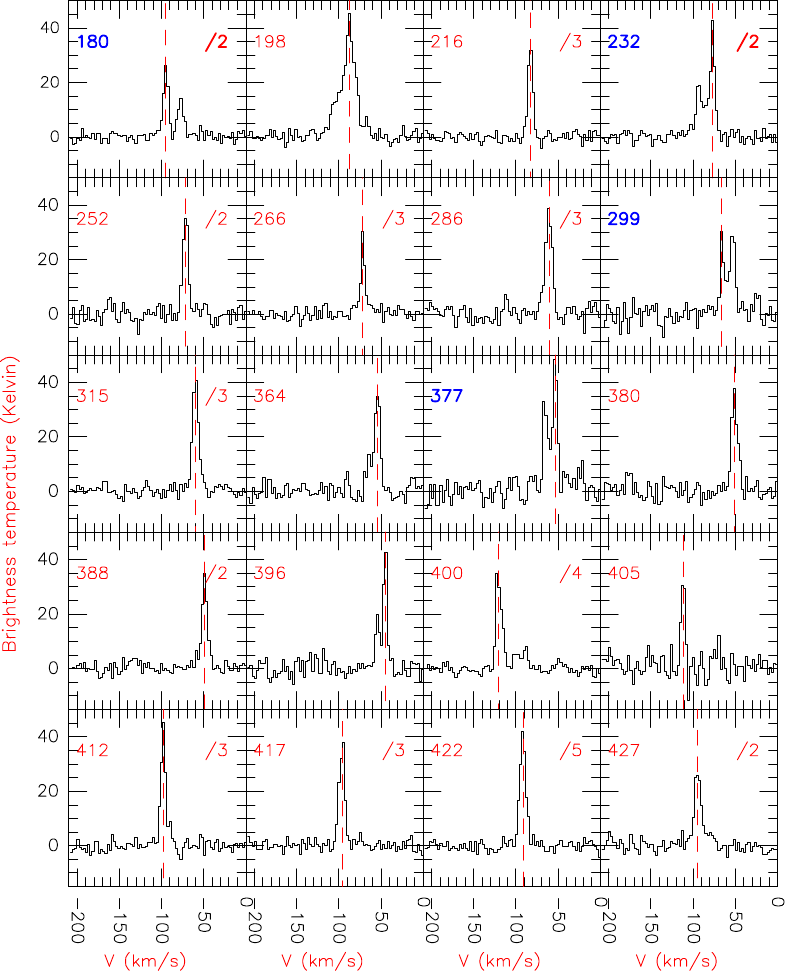}
		\contcaption{}
	\end{figure*}
	
	\subsection{Size versus velocity dispersion}\label{Larson_first}
	
\begin{figure}
		\centering
		\includegraphics[trim={0cm 0 0.5cm 0},clip,width=9cm]{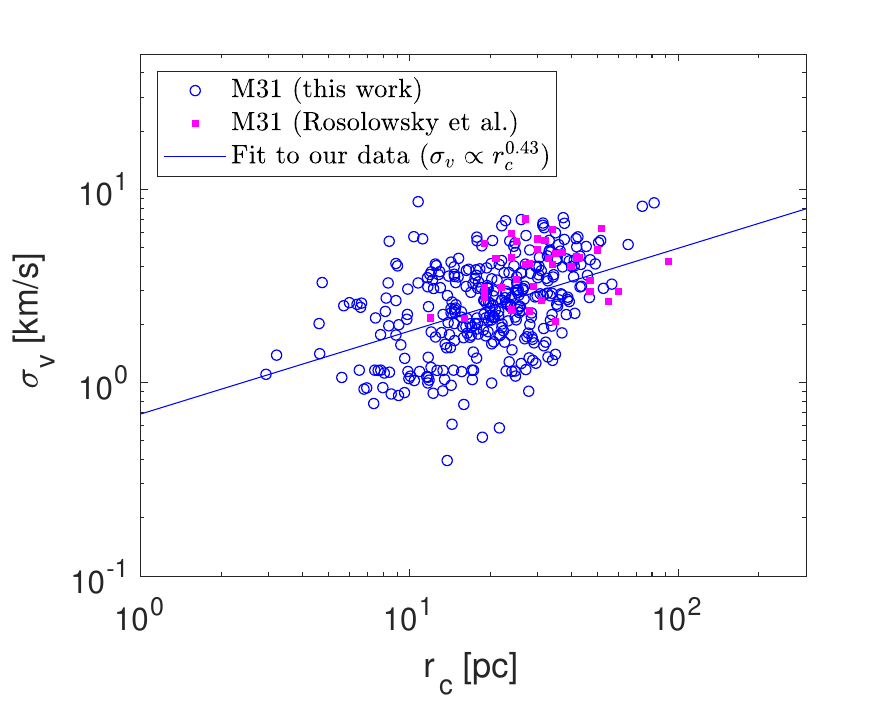}
		\caption{Source radius as a function of the $\sigma_{\rm v}^d$ parameter. The blue circles are the values obtained in this study, while the filled pink squares are values previously calculated for clouds of M31 \citep{2007ApJ...654..240R}. 
		The blue line is a fit to our data (see Section \ref{Larson_first}).}
		\label{radius_vs_sigma}
	\end{figure}

	We plotted values of $r_{\rm c}$ versus $\sigma_{\rm v}^d$ in Figure \ref{radius_vs_sigma}, where for comparison purposes we have also included the values derived by \cite{2007ApJ...654..240R} for M31 clouds observed with the BIMA interferometer.
	The BIMA data have a spatial resolution within 7.1-14.0 arcsec and a spectral resolution within 2.03-3.04 km s$^{-1}$. We fitted the relationship between the $r_{\rm c}$ versus $\sigma_{\rm v}^d$ for the M31 sources using the KMPFIT module of the Python package Kapteyn \citep{2016ascl.soft11010T}, finding $\sigma_{\rm v}=10^{-0.16\pm0.07}\,r_{\rm c}^{0.43\pm0.05}$ with the Pearson correlation coefficient of 0.43. 
	As seen in Figure \ref{radius_vs_sigma}, the $\sigma_{\rm v}^d$ and r$_{\rm c}$ values we derive for M31 are in agreement with those derived by \cite{2007ApJ...654..240R}. The slope of 0.43 found for the M31 clouds are in agreement with that for Milky Way (MW) clouds \citep{2014ApJS..212....2G,2016ApJ...822...52R} and also with that of 0.40$\pm$0.07 reported by \cite{2024ApJ...966..193L} for a sample of 117 clouds in M31.
	
	\subsection{Size versus mass}\label{Sizevsmass}
	
\begin{figure}
	\centering
	\includegraphics[trim={1.0cm 0 1cm 0},clip,width=9cm]{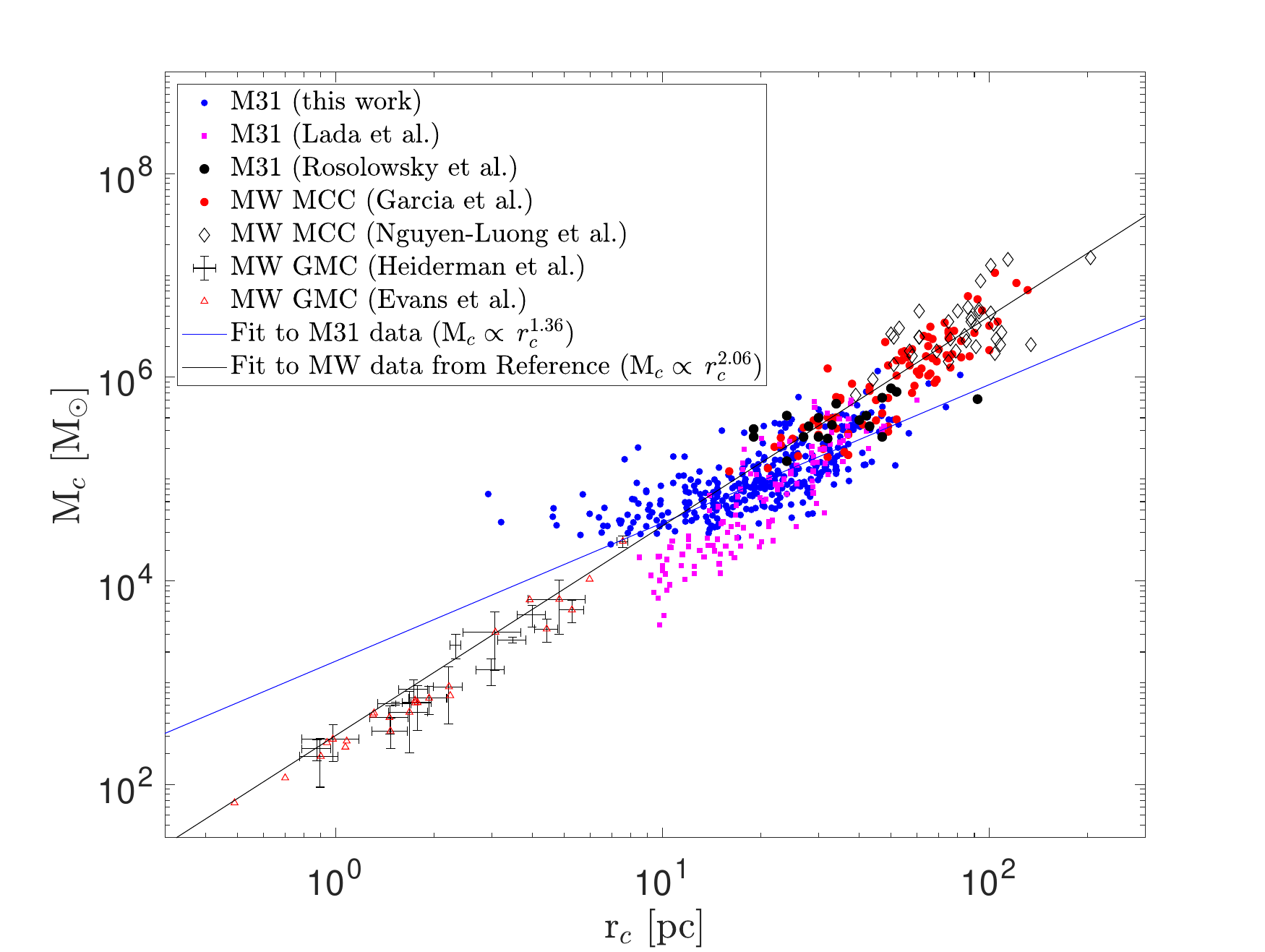}
	\caption{Radius versus mass for the clouds identified in M31. We also include here data derived previously for M31 clouds \citep{2007ApJ...654..240R,2024ApJ...966..193L}, MW MCCs \citep{2014ApJS..212....2G,2016ApJ...833...23N}, and MW clouds \citep{2010ApJ...723.1019H,2014ApJ...782..114E}.
	The blue line is a fit to our data together with data from reference \citep{2007ApJ...654..240R,2024ApJ...966..193L}, while the black line is a fit to MW MCCs (see Section \ref{Sizevsmass}).}
	\label{Radiusvsmass}
\end{figure}

	We have also plotted the radius versus the mass for our M31 clouds in Figure \ref{Radiusvsmass}. In this figure, we have also included data for MW molecular cloud complexes (MCC) studied by \cite{2014ApJS..212....2G} and \cite{2016ApJ...833...23N}. The MCC masses derived by both authors were determined considering a slightly different $X$ factor than the one used in our study, thus their masses were rescaled considering our $X$ factor.
	Figure \ref{Radiusvsmass} reveals that the radii and masses of our M31 clouds are lower than those of MW MCCs \citep{2014ApJS..212....2G,2016ApJ...833...23N}. 
	We have also included data for M31 clouds, as calculated by \cite{2007ApJ...654..240R} and \cite{2024ApJ...966..193L}, in Figure \ref{Radiusvsmass}. The comparison shows that our mass measurements are greater than those reported by \cite{2024ApJ...966..193L} for radii between $\sim$10 and $\sim$20 pc. However, our mass estimates align with those of \cite{2024ApJ...966..193L} for radii larger than $\sim$20 pc. In addition, our data matches the findings from \cite{2007ApJ...654..240R}, as seen in Figure \ref{Radiusvsmass}.
	The masses of giant molecular clouds (GMCs) in the MW, as determined by \cite{2010ApJ...723.1019H} and \cite{2014ApJ...782..114E} using extinction maps, are also included in Figure \ref{Radiusvsmass}. Their sources show smaller masses and radii than those of our M31 clouds.
	
	Using the KMPFIT module, we fitted our M31 data together with the data from \cite{2007ApJ...654..240R} and \cite{2024ApJ...966..193L} given in Figure \ref{Radiusvsmass}, finding the best-fit parameters $a$ and $N$ together with the correlation coefficient (C$_{\rm corr}$) given in Table \ref{sizemass}.
	Fitting the data for MW MCCs from \cite{2014ApJS..212....2G} and \cite{2016ApJ...833...23N}, we found the parameters $a$ and $N$ listed also in Table \ref{sizemass}. The slope $a$ of 2.06 is stepper than that found for the clouds in M31, which implies that the mass of the clouds studied in M31 does not scale with radius in the same way as in the MW MCCs.
	In our analysis, we determined a slope of 1.06$\pm$0.05 when fitting only our data. This slope is shallower than the 1.36 obtained by fitting all the data for M31 included in Figure \ref{Radiusvsmass}. The slope of 1.36 is also shallower than the 1.9 slope derived for MW GMCs \citep{2016ApJ...833...23N}.

\begin{table}
	\centering
	\caption{Best-fit parameters for the $M_{\rm c}=a\,r_{\rm c}^N$ relationship.}
	\label{sizemass}
	\begin{tabular}{lccr} 
		\hline
		Data set & a & N  & C$_{\rm corr}$\\
		\hline
		Our data + data from Lada et al. & 10$^{3.21\pm0.08}$ & 1.36$\pm$0.06 & 0.73 \\
		and Rosolowsky et al.  &                    &                &     \\                                                         
		MW data of MCCs  & 10$^{2.48\pm0.19}$ & 2.06$\pm$0.10  & 0.87\\
		\hline
	\end{tabular}
\end{table}

\subsection{Star formation}\label{KSlaw}

\begin{figure}
	\centering
	\includegraphics[width=9cm]{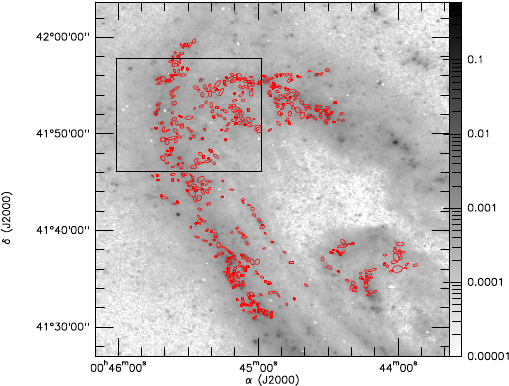}
	\caption{Surface density map of star formation in units of M$_\odot$ yr$^{-1}$ kpc$^{-2}$ from a previous work \citep{2013ApJ...769...55F}. Red ellipses (having the exact area of the structure) were determined from the best-fits to the sources identified by our dendrogram analysis. The black rectangle is a region zoomed in Figure \ref{Zoom_clouds}.}
	\label{SFRmap}
\end{figure}

\begin{figure}
	\centering
	\includegraphics[trim={0 4.5cm 0 4.5cm},clip,width=9cm]{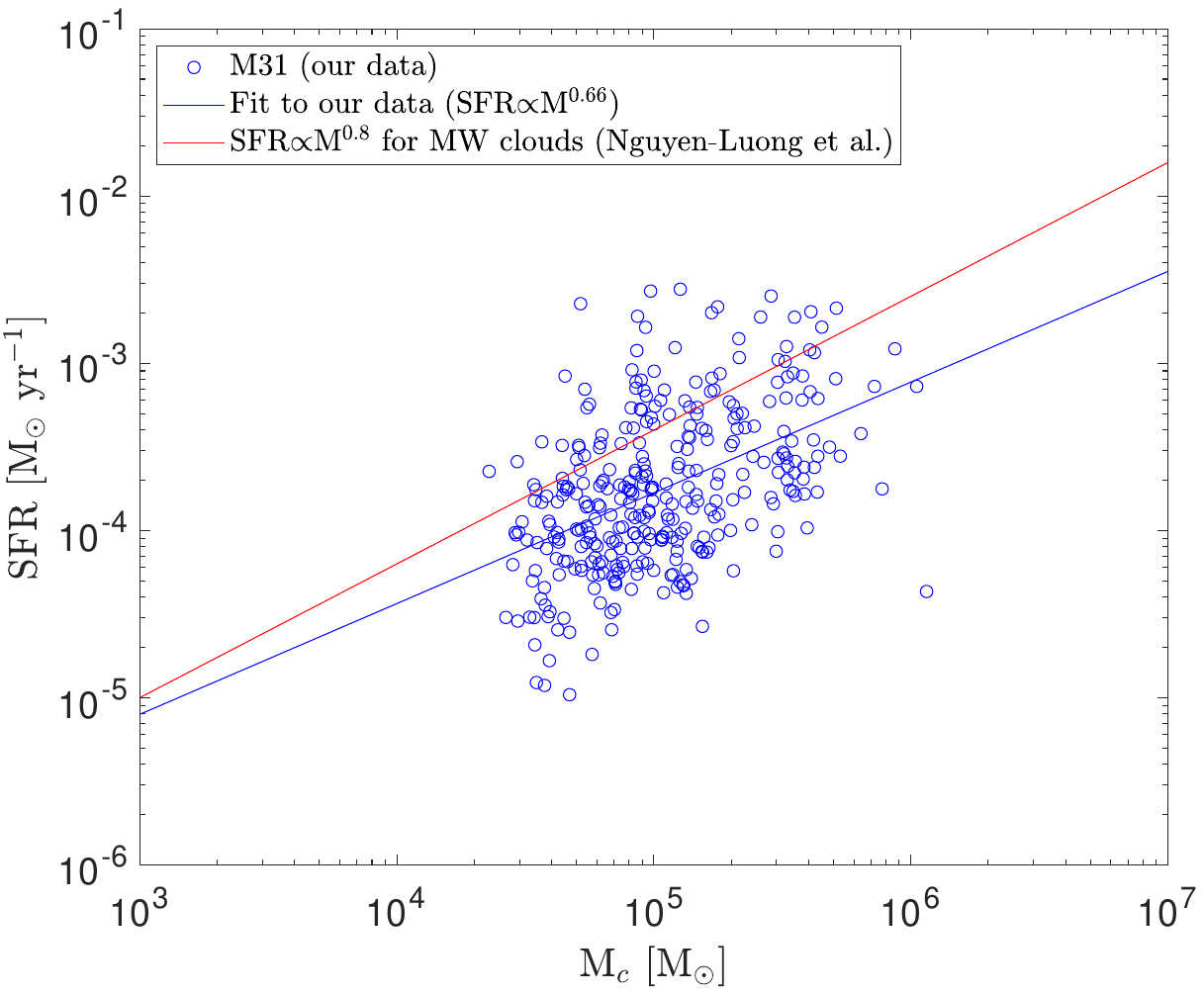}
	\caption{M$_{\rm c}$ versus the SFR for the clouds identified in M31. The blue line is the fit to our M31 data, while the red line is a fit to data of MW clouds obtained previously \citep{2016ApJ...833...23N}.}
	\label{SFRvsMass}
\end{figure}

\begin{figure}
	\centering
	\includegraphics[width=9cm]{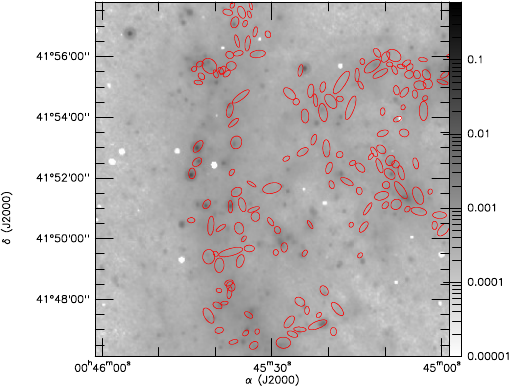}
	\caption{A zoomed region of the surface density map of star formation in units of M$_{\odot}$ yr$^{-1}$ kpc$^{-2}$ shown in Figure \ref{SFRmap}. The ellipses in red are best-fits to CO J=1-0 clouds obtained with our dendrogram analysis.}
	\label{Zoom_clouds}
\end{figure}

We used the surface density map of star formation (see Figure \ref{SFRmap}) obtained by \cite{2013ApJ...769...55F} to infer the star formation rate (SFR) values for the clouds identified in our study, which are plotted as a function of the source mass in Figure \ref{SFRvsMass}. In these plots and our analysis we have included only the 453 sources with one velocity component identified in our dendrogram analysis (Section \ref{COspectra_sec}). 
This is necessary since the contribution of the SFR to each source is unknown when there is an overlap of sources along the line of sight.
For calculating the SFR values, we first extracted the values of the surface density of SFR in units of $M_{\odot}$ yr$^{-1}$ kpc$^{-2}$ from the exact areas of the sources identified in our dendrogram analysis, then multiplied each value by its exact source area in units of kpc$^2$.
The surface density map of SFR was derived using the FUV and 24 $\mu$m emission, and it has a FWHM beam-width of 6\arcsec and a pixel size of 1.5\arcsec \citep{2013ApJ...769...55F}, which are very close to those of our CARMA CO J=1-0 data.
The 453 red ellipses in Figure \ref{SFRmap} have the exact area of the sources extracted with the dendrogram approach. These ellipses follow well the boundaries of the source isosurfaces (see Figure \ref{Identified_clouds}).
A close-up of this figure is presented in Figure \ref{Zoom_clouds}, where we can see that the extracted clouds do not always coincide with the emission peaks of the SFR map. This is consistent with the findings that young star clusters in M31 with ages of $<$10$^{7.2}$ yr are closer to GMCs than older star clusters \citep{2023MNRAS.522.6137P}. The resolution of our observations is lower than the mean separation of 100-300 pc between GMCs and \ion{H}{II} regions \citep{2020MNRAS.493.2872C}, which allows us to see an offset between the emission peaks of molecular clouds and star-forming regions in M31. 
\cite{2010ApJ...722.1699S} found a scale dependence of the depletion time (H$_{\rm 2}$ mass over SFR) on the spatial resolution, ranging from 75 pc to 1.2 kpc in M33.
At large scales, the depletion time reflects a mean value of the sampled region, while at scales (75 pc) of individual objects, the depletion time depends on their evolutionary state. This dependence agrees with the offsets between the clouds and the emission peaks of the SFR map in Figure \ref{SFRmap}.
This dependence is also thought to be responsible for the increase in the slope dispersion in the Kennicutt-Schmidt law of nearby star-forming galaxies at a high spatial scale of 100 pc compared to lower spatial scales >1 kpc \citep{2021A&A...650A.134P}.

We find that there is a correlation between the SFR values and the $M_{\rm c}$ values in Figure \ref{SFRvsMass}, which we fit using the KMPFIT module. The best-fitting parameters of the $SFR=a\,M_{\rm c}^N$ relationship of our M31 data are given in Table \ref{Bestfits2}. 
Our slope of 0.66 supports the idea that the Kennicutt-Schmidt law is not superlinear on scales of $\sim$22 pc \citep{2013ApJ...769...55F} in M31. Our slope of 0.66 agrees within its error bars with that of 0.6 obtained by \cite{2013ApJ...769...55F} based on a pixel by pixel basis, which is different from the methodology used in our work to study the Kennicutt-Schmidt law. 
The slope of 0.66 found for our M31 clouds is slightly lower than those of $\sim$0.8 derived for MW clouds \citep{2012ApJ...745..190L,2016ApJ...833...23N}. 
This can be seen when we compare our best fit with that for MW clouds (refer to Figure \ref{SFRvsMass}).

Our best fit in Figure \ref{SFRvsMass} shows SFR values for M31 that are slightly lower than those for Milky Way clouds by a factor of 2.5 for cloud masses of 10$^5$ M$_\odot$. However, this comparison must be taken with some caution, as the SFR values for the local clouds were estimated through counting of YSO (young stellar objects) with ages of $\sim$2 Myr \citep{2025A&A...693A..51D}, which differs from the methodology applied to calculate the SFR values for M31. The surface density map of SFR for M31 was derived from FUV emission and 24 $\mu$m emission, which are sensitive to star formation timescales of $\sim$5-100 Myr \citep{2012ARA&A..50..531K}. Our comparison of the SFR values between M31 and the Milky Way assumes that the SFR in M31 has been constant over a period of $\sim$100 Myr. Another caveat regarding our SFR values is that the surface density map of SFR used in our study was derived through a linear combination of FUV and 24 $\mu$m emission based on calibrations for entire galaxies \citep{2008AJ....136.2782L,2013ApJ...769...55F}. This method may not be applicable at the cloud scales in M31. Fortunately, the CO and SFR data used in our study have high spatial resolutions, which enables an effective comparison of our cloud masses and SFR values with those of Milky Way clouds.

\begin{figure}
	\centering
	\includegraphics[trim={0 0 0 0},clip,width=9cm]{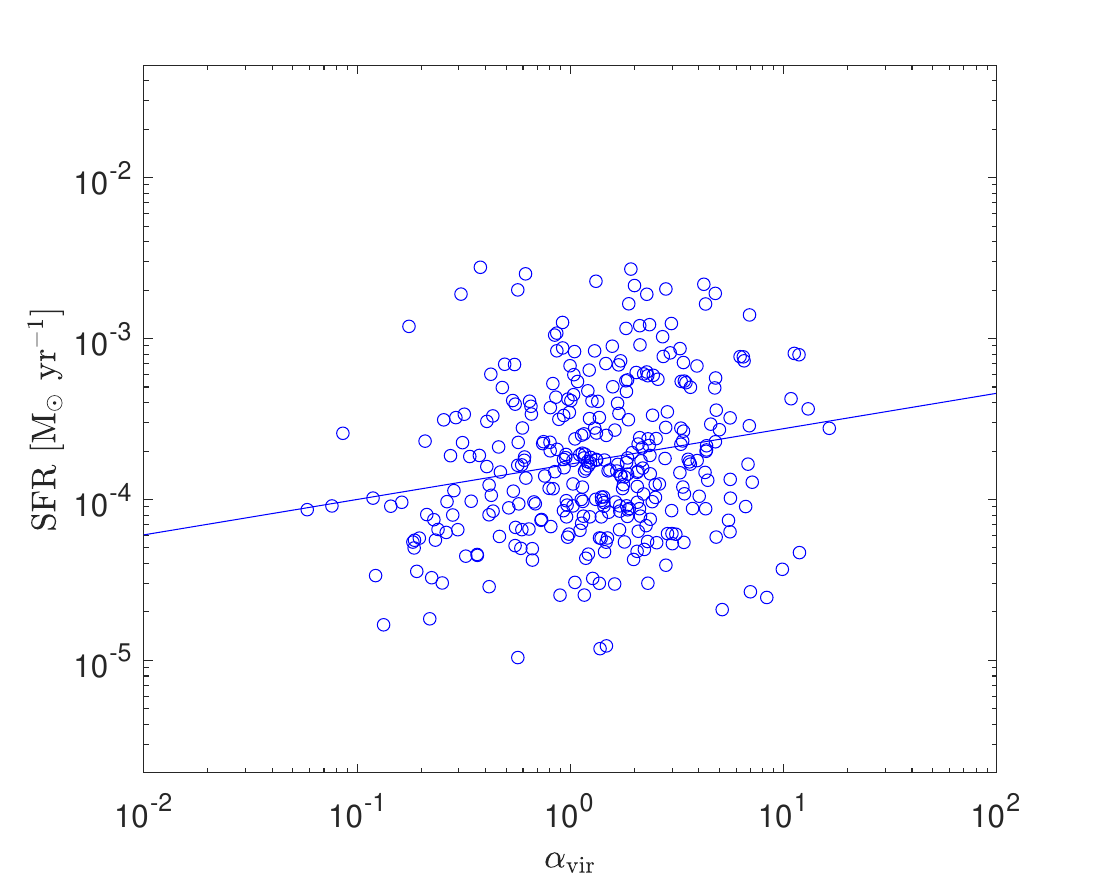}
	\caption{The $\alpha_{\rm vir}$ ratios represented as a function of the SFR values for our sample of clouds. The blued line shows the best fit to our data.}
	\label{SFRalpha}
\end{figure}

In addition, we find a weak dependence between the SFR and the virial ratio (shown in Figure \ref{SFRalpha}), which was expected due to the weak dependence between the $M_{\rm c}$ and $\alpha_{\rm vir}$ as well as the dependence between the SFR and $M_{\rm c}$. The best-fit parameters for our data given in Figure \ref{SFRalpha} are listed in Table \ref{alpha_SFR}.

\begin{table}
	\centering
	\caption{Best fit parameters for the SFR=$a\,M_{\rm c}^N$ relationship.}
	\label{Bestfits2}
	\begin{tabular}{llccr} 
		\hline
		Data & a & N  & C$_{\rm corr}$ \\
		\hline
		Our data & 10$^{-7.09\pm0.34}$ & 0.66$\pm$0.07 & 0.47\\
		\hline
	\end{tabular}
\end{table}

\begin{table}
	\centering
	\caption{Best fit parameters for the SFR=$a\,\alpha_{\rm vir}^N$ relationship.}
	\label{alpha_SFR}
	\begin{tabular}{lccrr} 
		\hline
		Data & a & N  & C$_{\rm corr}$ & p-value\\
		\hline
		Our data & 10$^{-3.78\pm0.03}$ & 0.22$\pm$0.06 & 0.20 & 3.4$\times$10$^{-4}$\\
		\hline
	\end{tabular}
\end{table}

\section{Conclusions}\label{Conclusions}

We identified 453 clouds in M31 by computing a dendrogram to CO J=1-0 data observed with CARMA in the position-position-velocity space. In addition, we identified 35 sources that show multiple velocity components, which are considered cloud complexes.
Using the information obtained from the dendrogram analysis, we calculated the radius ($r_{\rm c}$), velocity dispersion ($\sigma_{\rm v}^d$), CO-based mass ($M_{\rm c}$) and virial mass ($M_{\rm vir}$) for the clouds and cloud complexes in M31.
The cloud catalogue presented here is the largest for M31 so far.
In addition, we examined the relationships between size and velocity dispersion, size and mass, as well as star formation rate and mass for the clouds in M31. 
The main conclusions of our study are as follows:

\begin{enumerate}
	\item For the clouds in M31 we found mean values of 2.8 km s$^{-1}$, 22.1 pc, and 5.2 M$_{\odot}$ for the $\sigma_{\rm v}^d$, $r_{\rm c}$, and $\log_{10}(M_{\rm c})$, respectively. We did not find a correlation of the galactocentric radius with $r_{\rm c}$ or $\log_{10}(M_{\rm c})$. On the other hand, we discovered a weak anti-correlation between the galactocentric radius with the $\sigma_{\rm v}^d$ value. The clouds in M31 reveal values of 2.0 and 1.4 for the mean and median, respectively, of their virial parameters. Our findings indicate that 66\% of the clouds in M31 appear to be gravitationally bound. 
	Additionally, our analysis of the relationship between $\frac{\sigma_{\rm v}^2}{r_{\rm c}}$ and the surface density ($\Sigma_{\rm c}$) supports the idea that the majority of the objects studied in M31 are indeed bound.
	
	\item We carried out nonlinear least-squares fitting of the size-velocity dispersion relationship of our data using the KMPFIT module of Python, finding a slope of 0.43$\pm$0.05, which agrees with those found previously for clouds of the Milky Way and M31. 
	
	\item We compared the size and mass values of our clouds with those of the Milky Way and other M31 clouds. We found a slope of 1.36$\pm$0.06 when fitting our data together with reference data for M31. This slope is shallower than the 2.06 value obtained for molecular cloud complexes of the Milky Way. In addition, our slope of 1.36 is also shallower than that of 1.9 found for clouds in the Milky Way \citep{2016ApJ...833...23N}. This result shows that the mass of our studied clouds in M31 does not scale with radius in the same way as in clouds and cloud complexes of the Milky Way. The slope of 1.36 is steeper than the 1.06$\pm$0.05 calculated when fitting only our M31 data.
	
	\item Finally, we found that the ellipses that best fit the isosurfaces of the identified clouds in M31 do not always coincide with emission peaks of a surface density map of SFR, which was somewhat expected since the mean separation of GMCs and \ion{H}{II} regions is 100-300 pc \citep{2020MNRAS.493.2872C}, much larger than the spatial resolution of our CARMA observations. These offsets are in agreement with previous findings about the dependence of the depletion time (H$_{\rm 2}$ mass/SFR) on the evolutionary state of individual objects at pc scales. We found a slope of 0.66$\pm$0.07 for the Kennicutt-Schmidt law. This slope indicates that the Kennicutt-Schmidt law is not superlinear at scales of approximately 22 pc in M31. Additionally, our slope agrees with that calculated by \cite{2013ApJ...769...55F} despite the difference in the methodologies used to study the Kennicutt-Schmidt law.
    Our slope of 0.66 is slightly lower than those previously derived for clouds in the Milky Way.
	
\end{enumerate}





\section*{Acknowledgments}

This research made use of astropy \citep{2013A&A...558A..33A}, matplotlib\footnote{https://matplotlib.org/}, python packages and iPython \citep{2007CSE.....9c..21P}. We also use the KMPFIT module of the Python package Kapteyn \citep{2016ascl.soft11010T}, MIRIAD \citep{1995ASPC...77..433S}, GILDAS\footnote{https://www.iram.fr/IRAMFR/GILDAS/}, and MATLAB\footnote{https://www.mathworks.com/products/matlab.html}. The authors thank the anonymous referee for their helpful comments that improved the quality of the manuscript.


\section*{Data Availability}

The CO J=1-0 data cube and SFR map used for this work will be shared on reasonable request to the corresponding author.




\bibliographystyle{mnras}
\bibliography{bibliografia} 




\appendix

\section{Source Catalogue}

\begin{table*}
	\centering
	\caption{Properties of six clouds and four cloud complexes extracted from the CO J=1-0 data cube using a dendrogram. The complete list of 453 clouds and 35 cloud complexes is available as supplementary material.}
	\begin{threeparttable}
		\begin{tabular}{cccccccc}
			\hline
			Source number  & RA [J2000] & Dec [J2000]  &  r$_{\rm c}$\tnote{b} & $\sigma_{\rm v}$\tnote{b} & M$_{\rm c}$\tnote{b} & M$_{\rm vir}$\tnote{c} & SFR\tnote{d} \\
			              & ($^\circ$) & ($^\circ$)   & (pc)         & (km s$^{-1}$) & ($\times$10$^4$ M$_{\odot}$) & ($\times$10$^4$ M$_{\odot}$) & ($\times$10$^{-4}$ M$_\odot$ yr$^{-1}$) \\
			\hline
			1            & 11.2243 & 41.5261 & 5.61           & 1.06           & 2.82           & 0.66           & 0.624 $\pm$0.013   \\
			\hline
			2            & 11.2475 & 41.5274 & ...            & 2.12$\pm$2.52  & 4.10$\pm$1.05  & ...            & 2.461 $\pm$ 0.015   \\
			\hline
			3            & 11.2553 & 41.5297 & 13.94$\pm$4.89 & 2.04$\pm$0.61  & 3.21$\pm$1.84  & 6.05$\pm$4.18  & 0.876 $\pm$ 0.021   \\
			\hline
			4            & 11.2718 & 41.5975 & ...            & ...            & 2.16$\pm$3.22  & ...            & 0.287 $\pm$ 0.012    \\
			\hline
			5            & 11.2599 & 41.5397 & 20.60$\pm$9.17 & 2.98$\pm$2.36  & 8.78$\pm$2.85  & 19.06$\pm$31.37& 3.338 $\pm$ 0.060    \\
			\hline
			6\tnote{a}   & 11.2715 & 41.5392 & 19.32 & ...            & 4.04$\pm$6.30  & ...            & ...     \\
			\hline
			7\tnote{a}   & 11.2575 & 41.5448 & 19.40$\pm$9.14 & 3.65$\pm$2.41  & 9.09$\pm$2.83  & 26.81$\pm$37.58& ...     \\
			\hline
			8\tnote{a}   & 11.2757 & 41.5498 & ...            & 0.86  & 2.69$\pm$1.74  & ...            & ...      \\
			\hline
			9            & 11.2635 & 41.5521 & 16.94$\pm$4.95 & 2.93$\pm$1.09  & 6.79$\pm$1.58  & 15.17$\pm$12.11& 1.789 $\pm$ 0.042    \\
			\hline
			10\tnote{a}  & 11.3001 & 41.5842 & 6.95           & 1.01           & 4.57           & 0.74           & ...                 \\
			\hline
			
		\end{tabular}
		\begin{tablenotes}
			\item[a] A cloud complex with multiple velocity components identified in our dendrogram analysis.
			\item[b] Our dendrogram analysis does not provide the values for $r_{\rm c}$ and/or $\sigma_{\rm v}$ for some sources. In this table, $\sigma_{\rm v}$ and r$_{\rm c}$ are deconvolved values. In some cases, there are no errors listed for these columns because the bootstrap method (see Section 3.2) does not yield errors for certain parameters used in error propagation calculations.
			\item[c] We cannot calculate the $M_{\rm vir}$ value for some sources due to the unavailability of their $\sigma_{\rm v}$ and/or r$_{\rm c}$ values. We also are unable to derive the error for the $M_{\rm vir}$ value for some sources because the error for $\sigma_{\rm v}$ and/or r$_{\rm c}$ are/is not available.
			\item[d] 35 cloud complexes that do not have an SFR value due to their multiple velocity components along the line of sight, which complicates the inference of the SFR contribution from each component.
		\end{tablenotes}
	\end{threeparttable}\label{cloud_catalogue}
\end{table*}



\bsp	
\label{lastpage}
\end{document}